\documentclass[onecolumn]{article}
\usepackage{amsmath,amssymb,graphicx,units}
\usepackage[usenames,dvipsnames]{color}
\usepackage{lipsum}
\usepackage{setspace}
\usepackage{ulem}
\usepackage[bookmarks=false,linkcolor=blue,urlcolor=blue,colorlinks,citecolor=blue]{hyperref}

\usepackage{mathrsfs}
\usepackage[a4paper,left=2.5cm,right=2.5cm, top=1.0cm,bottom=1.5cm,textwidth=20cm,textheight=20.3cm]{geometry}

\begin{document}
\date{}
\title{An analytical variational method for the biased quantum Rabi model in the ultra-strong coupling regime\thanks{Project supported by National Natural Science Foundation of China (Grant No. 11674139, No. 11604009, No. 11704025), Supported by Program for Changjiang Scholars and Innovative Research Team in University (Grant No. IRT-16R35) and the Fundamental Research Funds for the Central Universities. ZJY also acknowledges the financial support of the Future and Emerging Technologies (FET) programme within the Seventh Framework Programme for Research of the European Commission, under FET-Open Grant No. 618083 (CNTQC).}}


\author{Bin-Bin Mao$^{1}$, \ Maoxin Liu$^{2}$, \ Wei Wu$^{2}$,\\ \ Liangsheng Li$^{3}$, \ Zu-Jian Ying$^{4}$, \ Hong-Gang Luo$^{1,2}$\thanks{Corresponding author. E-mail:luohg@lzu.edu.cn}\\
$^{1}${Center of Interdisciplinary Studies $\&$ Key Laboratory for Magnetism and Magnetic Materials}\\{ of the Ministry of Education, Lanzhou
	University, Lanzhou 730000, China}\\  
$^{2}${Beijing Computational Science Research Center, Beijing 100193, China}\\ 
$^{3}${Science and Technology on Electromagnetic Scattering Laboratory, Beijing 100854, China}\\   
$^{4}${CNR-SPIN, I-84084 Fisciano (Salerno), Italy and Dipartimento di Fisica ``E. R. Caianiello,"}\\{ Universit$\mathrm{\grave{a}}$ di Salerno, I-84084 Fisciano (Salerno), Italy}}   



\maketitle

\begin{abstract}
An analytical variational method for the ground state of the  biased quantum Rabi model in the ultra-strong coupling regime is presented. This analytical variational method can be obtained by a unitary transformation or alternatively by assuming the form of ground state wave function. The key point of the method is to introduce a variational parameter $\lambda$, which can be determined by minimizing the energy functional. Using this method, we calculate  physical observables with high accuracy in comparison with the numerical exact one. Our method evidently improves the results from the widely used general rotating-wave approximation (GRWA) in both qualitative and quantitative ways.
\end{abstract}



\section{Introduction}
The Quantum Rabi model \cite{rabi1936,rabi1937} plays a fundamental role in many fields of modern physics, such as quantum optics \cite{Walther2006}, quantum information \cite{raimond2001}, and condensed matter physics \cite{Holstein1959}. It has been used to characterize a system composed of a two-level atom and a single mode bosonic field with a linear coupling in between,
namely, the so-called cavity quantum electromagnetic dynamics (QED) system \cite{Walther2006}. Recently, the model has also been realized in many other promising quantum systems, e.g. the superconducting circuits system \cite{Wallraff2004,wallraff2005}, trapped ion \cite{puebla2017} and so on. A remarkable experimental achievement involves the realization of the so-called ultra-strong coupling \cite{Fornd2017,Yoshihara2017,chen2017}. The realization from strong coupling to ultra-strong coupling and to even the deep-strong coupling inspires theorist to explore the interesting physics involved in the model. In fact, the quantum Rabi model has been extensively studied theoretically and a series of important features have been found, e.g. its integrability \cite{braak2011}, its geometric phase \cite{larson2012,Wang2012_cpb}, the polaron physics \cite{ying2015,liu2017jpa,cong2017} . In particular, more recently, quantum phase transitions and its universality class have been found in such a finite degree of freedom system \cite{hwang2015,hwang2016,puebla2017,liu2017arxiv} .

The starting point of the theoretical research on such a quantum model lies in solving the eigenvalue problem of it. Although the form of the model looks quite simple, it is not easy to solve it exactly. In fact, it is only up to 2011 that the model has been exactly solved by Braak \cite{braak2011}. This pioneering work triggered a series of exact solution in related models \cite{xie2014,Wolf2012,Chen2012,Batchelor2015,Batchelor2016}. However, since the key technique of the exact solution involves solving the zero points of a transcendental function, it is not trivial to use it to analyze the physical properties of the model.  Hence, people are still trying to find some intuitive analytic approximation to calculate and understand the model. The earliest and might be the most famous approximation is the so-called rotating wave approximation (RWA). However, this approximation is only applicable to weak coupling with near-resonance condition, and thus it fails to satisfy the requirement of the experiment in current ultra-strong coupling regime\cite{Forn2010}. In 2007, Irish proposed the so-called general rotating wave approximation (GRWA) \cite{irish2007}, which does succeed in presenting the correct structure of the energy spectrum over a large parameter regime. It also provides a framework to analytically treat many models which are cousins of the quantum Rabi model \cite{yuanweizhang2011,lixianyu2012,zhang2013,Zhang2013_cpl,He2013_cpl,Dong2016_cpb,liu2015,zhang2015,zhang2016,zhang2016three,Liu2013,Wang2012,Gan2010,Yan2015,Mao2016}.

We notice that the so-called bias term is usually involved in the experimental circuit QED system\cite{Forn2010}. Therefore, the study of the Rabi model with the bias term is indispensable. However, most of theoretical studies on the quantum Rabi model focus on the unbiased case. Although there already exists an analytical discussion based on the GRWA for the biased quantum Rabi model\cite{zhang2013}, we found that there is still space to improve the accuracy of it in ultra-strong coupling regime, especially when the resonator frequency is small. Moreover, the ground state mean photon number and correlation predicted by GRWA are independent of the two level energy splitting and the bias strength. It is actually not true. For these reasons, we show in this work an improvement to the GRWA by introducing an analytical variational method. Our method can be obtained by two equivalent ways, namely, the unitary transformation method and the trial ground state wavefunction method. The key point is that we determine the optimal value (by minimizing the ground state energy) of a variational parameter $\lambda$ introduced in our method. To show the validity of our method and its improvement on the GRWA, we compare our results to the numerical exact ones and the ones from GRWA, respectively. As an approximation, we also discuss the applicable condition for the method we presented in the current work.

\section{Model Hamiltonian}
\label{sec_2}
The Hamiltonian of the biased quantum Rabi model reads\cite{Zhong2014,Xie2017,Li2015,Li2016}
\begin{equation}\label{eq:hamiltonian_0}
H_{B}=\omega a^{\dagger}a+g\sigma_{x}\left(a+a^{\dagger}\right)+\Delta\sigma_{z}+\epsilon^{\prime}\sigma_{x},
\end{equation}
where $a^{\dagger}$ ($a$) is the creation (annihilation) operators of the bosonic oscillator with frequency $\omega$, Pauli matrices $\sigma_{z}$ describes the two level system with energy splitting $2\Delta$, $\epsilon^{\prime}\sigma_{x}$ is the biased term and $g$ is the coupling strength.  Taking a rotation on $y$-axis of the spin and replacing the parameters as $\epsilon^{\prime}=\epsilon/2$ and $\Delta=\Omega/2$, we can obtain
\begin{eqnarray}\label{eq:hamiltonian}
H&=&e^{i\frac{\pi}{4}\sigma_{y}}H_{B}e^{-i\frac{\pi}{4}\sigma_{y}}\nonumber\\
&=&\omega a^{\dagger}a-\frac{\Omega}{2}\sigma_{x}+\frac{\epsilon}{2}\sigma_{z}+g\sigma_{z}\left(a^{\dagger}+a\right).
\end{eqnarray}

\section{Variational method}\label{sec_3}
Our main idea of the variational method is to introduce an unfixed parameter $\lambda$ in a unitary transformation (see below). The optimal value of $\lambda$ can be determined by minimizing the ground state energy. There are two ways to develop the variational scheme. One is to perform a $\lambda$-dependent unitary transformation, the other is to directly introduce a $\lambda$-dependent ground state wave function. Both these two ways are proved to be equivalent to each other.
\subsection{Variational method approached by a unitary transformation}
We start our variational scheme by performing a unitary transformation as similar to the GRWA \cite{irish2007,zhang2013}. Thus the Hamiltonian in Eq.~\eqref{eq:hamiltonian} is transformed into $\tilde{H}=UHU^{\dagger}$, where $U=e^{\lambda\sigma_{z}\left(a^{\dagger}-a\right)}$.
Here $\lambda$ is introduced as a variational parameter(as in Ref. \cite{liu2015}), which is in contrast to the GRWA where it is fixed to $\lambda = g/\omega$.
We will see later that this difference allows us to improve the GRWA. The transformed Hamiltonian is explicitly written as
\begin{equation}\label{eq:hamiltonian_u}
\tilde{H}=\tilde{H}_{0}+\tilde{H}_{1}+\tilde{H}_{2}+\tilde{H}_{3},
\end{equation}
where
\begin{eqnarray}
\tilde{H}_{0}&=&\omega a^{\dagger}a+\frac{\epsilon}{2}\sigma_{z}+\left(\omega\lambda^{2}-2\lambda g\right),\\
\tilde{H}_{1}&=&-\frac{\Omega}{2}\sigma_{x}{\rm cosh}\left[2\lambda\left(a^{\dagger}-a\right)\right],\\
\tilde{H}_{2}&=&-\frac{\Omega}{2}\left\{ {\rm i}\sigma_{y}{\rm sinh}\left[2\lambda\left(a^{\dagger}-a\right)\right]\right\}, \\
\tilde{H}_{3}&=&\left(g-\omega\lambda\right)\sigma_{z}\left(a^{\dagger}+a\right).
\end{eqnarray}

The hyperbolic sine and cosine terms can be further expanded as\cite{Irish2005}
\begin{eqnarray}
& &\mathrm{sinh}\left[\lambda\left({a}^{\dagger}-{a}\right)\right]\\
\!&=&\!\sum_{k=0}^{\infty}\left[\left({a}^{\dagger}\right)^{2k+1}F_{2k+1}\left({a}^{\dagger}{a}\right)+F_{2k+1}\left({a}^{\dagger}{a}\right){a}^{2k+1}\right],\nonumber
\end{eqnarray}
and
\begin{eqnarray}
\!& &\mathrm{cosh}\left[\lambda\left({a}^{\dagger}-{a}\right)\right]\\
\!&\!=\!&\!F_{0}\left({a}^{\dagger}{a}\right)\!+\!\sum_{k=1}^{\infty}\bigg[\left({a}^{\dagger}\right)^{2k}F_{2k}\left({a}^{\dagger}{a}\right)\!+\!F_{2k}\left({a}^{\dagger}{a}\right){a}^{2k}\bigg],\nonumber
\end{eqnarray}
respectively.
The function $F_{m}$ is defined as
\begin{equation}
F_{m}\left({n}\right)=e^{-\lambda^{2}/2}\lambda^{m}\frac{n!}{\left(n+m\right)!}L_{n}^{m}\left(\lambda^{2}\right),
\end{equation}
where $m$ and $n$ are integers, $L_{n}^{m}\left(x\right)=\sum_{i=0}^{n}\left(-x\right)^{i}\frac{\left(n+m\right)!}{\left(m+i\right)!\left(n-i\right)!i!}$ is the Laguerre polynomial.
Up to now, $\tilde{H}$ is exact, but it is still unsolvable. Therefore, we have to take some approximations to treat it analytically.

In this work, since we are interested in the low energy properties of the model, the photon excitations can be neglected in $\tilde{H}$. Thus we approaches to an effective approximate Hamiltonian:
\begin{equation}\label{heff}
\tilde{H}_{eff}=\omega a^{\dagger}a+\frac{\epsilon}{2}\sigma_{z}+\left(\omega\lambda^{2}-2g\lambda\right)-\frac{\Omega}{2} F_{0}\left(a^{\dagger}a\right)\sigma_{x}.
\end{equation}

The Hamiltonian in Eq.~\eqref{heff} is analytically solvable. We can choose the basis as $\left|\pm_z\right\rangle\otimes\left|n\right\rangle$, where the $\left|\pm_z\right\rangle$ are eigenstates of $\sigma_{z}$ with eigenvalues $\pm1$, and the $\left|n\right\rangle$ is the Fock state. The effective Hamiltonian can be readily divided into invariant subspace:
\begin{eqnarray}\label{eq:h0_matrix}
\tilde{H}_{n}\left(\lambda\right) & = & \left(\begin{array}{cc}
\xi_{n}^{-} & R_{n}\\
R_{n} & \xi_{n}^{+}
\end{array}\right),
\end{eqnarray}
where $\xi_{n}^{\pm}=\omega n\pm\epsilon/2+\omega\lambda^{2}-2g\lambda$,  $R_{n}=-\Omega F_{0}(n)/2$.

The eigenvalues of the matrix in Eq.~\eqref{eq:h0_matrix} can be  obtained as
\begin{equation}\label{eq:eigenvalues}
E_{n}^{\pm}\left(\lambda\right) =  \frac{\xi_{n}^{-}+\xi_{n}^{+}\pm\sqrt{\left(\xi_{n}^{-}-\xi_{n}^{+}\right)^{2}+4\left(R_{n}\right)^{2}}}{2},
\end{equation}
and the corresponding eigenvectors are
\begin{equation}\label{wf}
\left|\tilde{\phi}_{n}^{\pm}\left(\lambda\right)\right\rangle   =  \left( \alpha_{n}^{\pm}\left|-_z\right\rangle+\beta_{n}^{\pm}\left|+_z\right\rangle \right) \otimes |n\rangle,
\end{equation}
where
\begin{eqnarray}
\alpha_{n}^{\pm} & = & \pm\sqrt{\frac{1}{2}\left[1\pm\frac{\xi_{n}^{-}-\xi_{n}^{+}}{\sqrt{\left(\xi_{n}^{-}-\xi_{n}^{+}\right)^{2}+4\left(R_{n}\right)^{2}}}\right]},\\
\beta_{n}^{\pm} & = & \sqrt{\frac{1}{2}\left[1\mp\frac{\xi_{n}^{-}-\xi_{n}^{+}}{\sqrt{\left(\xi_{n}^{-}-\xi_{n}^{+}\right)^{2}+4\left(R_{n}\right)^{2}}}\right]}.
\end{eqnarray}

With the $\lambda$-dependent energy spectra and wave functions in hand, we can explore particularly the ground state properties and determine the optimal value of $\lambda$.  Specifically, the ground state energy is
\begin{equation}\label{eq:E_lambda}
E_{0}\left(\lambda\right)=\omega\lambda^{2}-2g\lambda-\frac{1}{2}\sqrt{\epsilon^{2}+\Omega^{2} e^{-4\lambda^{2}}},
\end{equation}
The $\lambda$ can be determined by minimizing the ground state energy $E_{0}(\lambda)$, namely, solving the equation
\begin{eqnarray}\label{eq:E_partial}
\frac{\partial E_{0}\left(\lambda\right)}{\partial\lambda}=0.
\end{eqnarray}
It can be solved in numerical way or approximately  solved as
\begin{eqnarray}\label{eq:lambda_0}
\lambda\approx\frac{g}{\omega+\Omega^{2}e^{-4{\lambda_0}^{2}}/\sqrt{\Omega^{2}e^{-4{\lambda_0}^{2}}+\epsilon^{2}}},
\end{eqnarray}
where $\lambda_0=g/\left(\omega+\Omega^{2}/\sqrt{\Omega^{2}+\epsilon^{2}}\right)$.

In the limit cases, we can further simplify the formula of $\lambda$. In the large $g$ limit, $\lambda$ approaches to $g/\omega$, which recovers the so-called GRWA \cite{irish2007,zhang2013}. In the small $g$ case, there is
\begin{equation}\label{lambda0}
\lambda =\frac{g}{\omega+\Omega^{2}/\sqrt{\Omega^{2}+\epsilon^{2}}},
\end{equation}
which is the same as $\lambda_0$. The small $g$ case involves the regime where $g$ is small but comparable to $\omega$. This limit covers the typical ultra-strong coupling regime, which recently attracts much experimental interests and is our main focus in this work.

Having determined the value of $\lambda$, we have fully obtained the analytical formula of the ground state wavefunction, and the physical observables can be thus calculated.

\subsection{Variational method approached by a trial ground state wavefunction}
Before going on, we present an alternative way to approach the variational scheme.
Instead of performing the unitary transformation, we can in fact directly  assume the form of trial ground state wave function. Remind that the trial wave function in
the unbiased Rabi model takes the form
\begin{equation}
|G\rangle_{unbias}=\frac{1}{\sqrt{2}}(|-_z\rangle\otimes|\lambda\rangle+|+_z\rangle\otimes|-\lambda\rangle).
\end{equation}
where $\left|\lambda\right\rangle=e^{-\left|\lambda\right|^{2}/2}\sum_{n=0}^{\infty}\lambda^{n}/\sqrt{n!}\left|n\right\rangle$ is a coherent state.
The equality of weight for up and down spin is due to the parity symmetry. Naturally, if the parity symmetry breaks down, the weights of up and down spins are no longer equal. Therefore, the trail ground state wave function can be assumed as
\begin{equation}\label{gs_wf}
\left|\phi_{0}\left(\lambda\right)\right\rangle =\alpha\left|-_z,\lambda\right\rangle+\beta\left|+_z,-\lambda\right\rangle.
\end{equation}
For normalization, we set $\alpha=\rm{cos}\left({\theta}/{2}\right)$ and $\beta=\rm{sin}\left({\theta}/{2}\right)$.
Thus the ground state energy should be
\begin{eqnarray}
E\left(\lambda\right)&=&\left\langle\phi_{0}\left(\lambda\right)\left|H\right|\phi_{0}\left(\lambda\right)\right\rangle\nonumber\\
& = & \omega\lambda^{2}-2g\lambda-\frac{1}{2}\left(\Omega e^{-2\lambda^{2}}{\rm sin}\theta+\epsilon{\rm cos}\theta\right)\nonumber\\
& \geq & \omega\lambda^{2}-2g\lambda-\frac{1}{2}\sqrt{\left(\Omega e^{-2\lambda^{2}}\right)^{2}+\epsilon^{2}}  .
\end{eqnarray}
In the last line, the equal condition is taken if and only if ${\rm sin}\theta={\Omega e^{-2\lambda^{2}}}/{\sqrt{\left(\Omega e^{-2\lambda^{2}}\right)^{2}+\epsilon^{2}}}$
and ${\rm cos}\theta={\epsilon}/{\sqrt{\left(\Omega e^{-2\lambda^{2}}\right)^{2}+\epsilon^{2}}}$.
Thus the ground state energy is
\begin{equation}\label{eq:gs_energy_lambda}
E_{0}\left(\lambda\right)=\omega\lambda^{2}-2g\lambda-\frac{1}{2}\sqrt{\epsilon^{2}+\Omega^{2} e^{-4\lambda^{2}}},
\end{equation}
which recovers Eq.~\eqref{eq:E_lambda}.
\begin{figure}[!tb]
	\centering	
	\includegraphics[width=0.9\columnwidth]{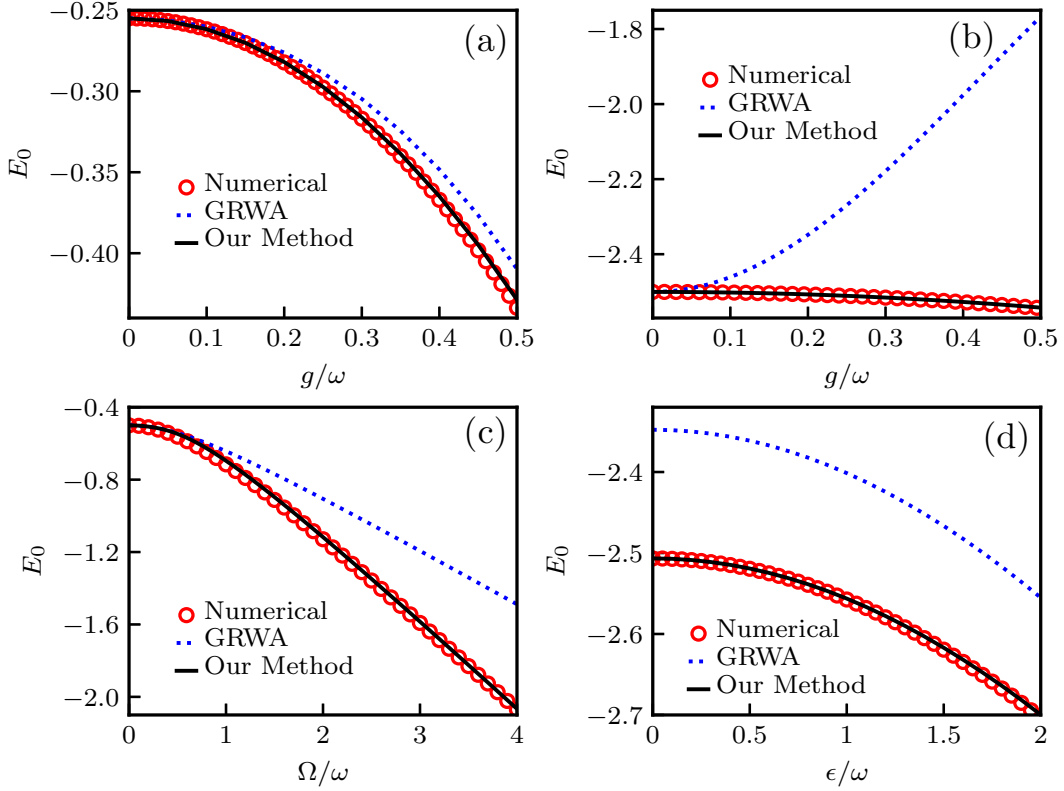}
	\caption{(color online) The ground state energy $E_{0}$ for different parameters, (a) $\epsilon=0.1$, $\Omega=0.5$, (b) $\epsilon=0.1$, $\Omega=5$, (c) $\epsilon=0.5$, $g=0.5$, (d) $g=0.2$, $\Omega=5$.}\label{fig1}
\end{figure}

\begin{figure}[!tb]
	\centering	
	\includegraphics[width=0.9\columnwidth]{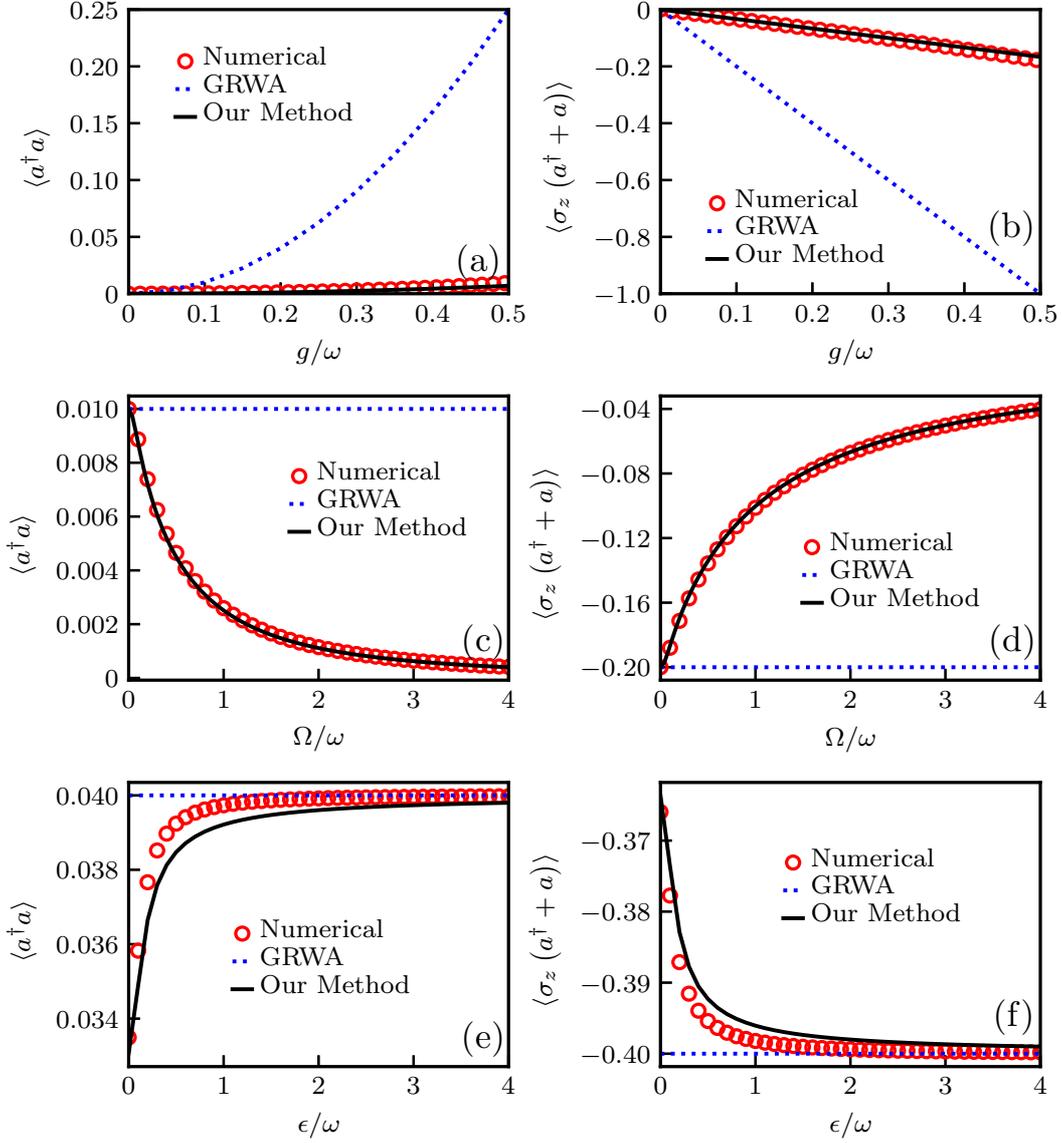}
	\caption{(color online)  The mean photon number $\left\langle a^{\dagger}a\right\rangle$ and coupling $\left\langle \sigma_{z}\left(a^{\dagger}+a\right)\right\rangle$ for different parameters, (a, b) $\epsilon=0.1$, $\Omega=5$, (c, d) $\epsilon=0.1$, $g=0.1$, (e, f) $g=0.2$, $\Omega=0.1$.}\label{fig2}
\end{figure}
\begin{figure}[!tb]
	\centering	
	\includegraphics[width=0.9\columnwidth]{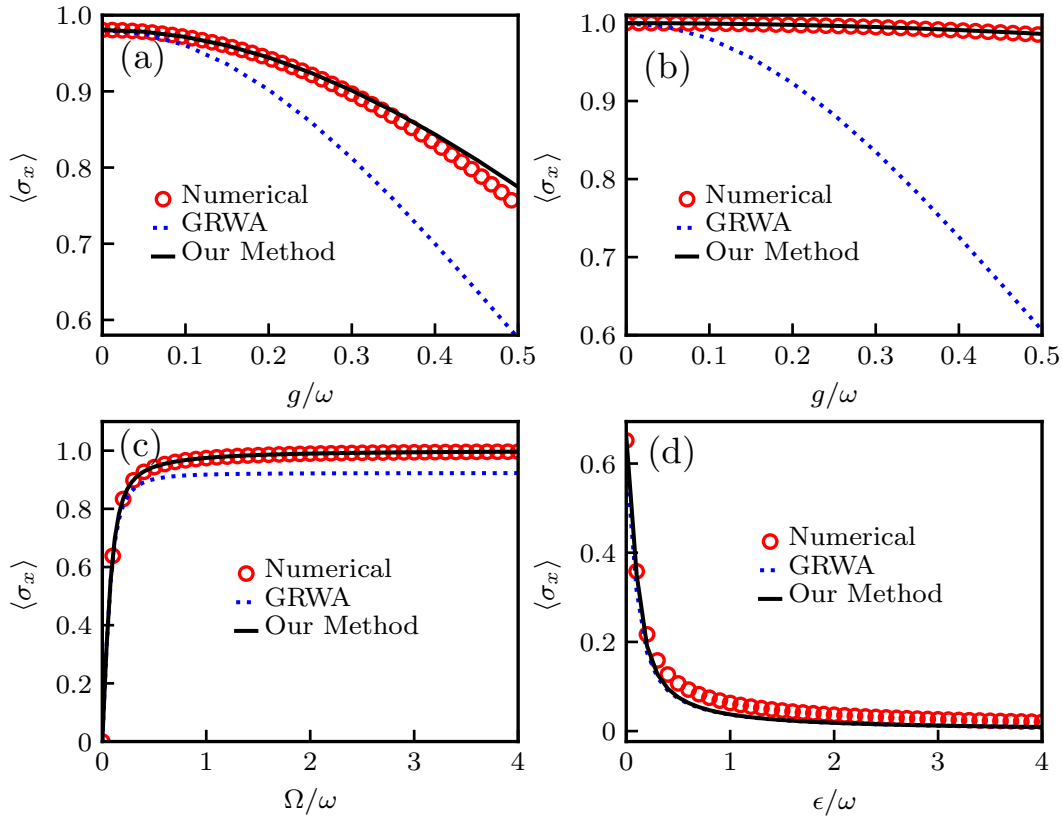}
	\caption{(color online)  The spin orientation $\sigma_{x}$ for different parameters, (a) $\epsilon=0.1$, $\Omega=0.5$, (b) $\epsilon=0.1$, $\Omega=5$, (c) $\epsilon=0.1$, $g=0.2$, (d) $g=0.5$, $\Omega=0.1$. }\label{fig3}
\end{figure}
\section{The results of the variational calculation comparing to the other methods}
\subsection{Calculations of the observables}
With the analytical expressed ground state wave function (Eq.~\eqref{wf} or Eq.~\eqref{gs_wf}) in hand, we can readily calculate the physical observables. Since we are interested in the ultra-strong coupling regime, the $\lambda$ is chosen as the analytical formula in Eq.~\eqref{lambda0}. The quantum average of a physical observable $Q$ is
$
\langle Q \rangle = \langle G| Q | G \rangle. 	
$
In this work, we take ground state energy $E_G=\langle H \rangle$, mean photon number $\langle a^{\dagger} a \rangle$, two level occupation $\langle \sigma_x \rangle$ and correlation $\langle \sigma_z(a+a^{\dagger}) \rangle$ as examples. Besides the ground state energy in Eq.~\eqref{eq:E_lambda}, the rest ones are:
\begin{eqnarray}
\left\langle a^{\dagger}a\right\rangle&=&\lambda^{2},\\
\left\langle \sigma_{z}\left(a^{\dagger}+a\right)\right\rangle&=&-2\lambda,
\end{eqnarray}
and
\begin{equation}
\left\langle \sigma_{x}\right\rangle=\frac{\Omega e^{-4\lambda^{2}}}{\sqrt{\epsilon^{2}+\Omega^{2}e^{-4\lambda^{2}}}}.
\end{equation}

We first consider the ground state energy. In Fig.~1, we see that our variational method agrees with the numerical exact one very well. Comparing to the result from GRWA, we find that our variational method shows a lower energy curve. It means that in spirit of variational way, the method successfully achieves the energy minimum. From Fig.~1, we can also obtain some detail properties of the variational method. In Fig.~1 (a) and (b), we see that the deviation between GRWA and numerical one becomes large when the coupling strength increases. It can be understood that $\lambda$ is proportional to $g$, and thus the increasing of $g$ enlarges the difference between the optimal $\lambda$ in our method and $\lambda=g/\omega$ in the GRWA. For the same reason, from Fig.~1 (c), we infer that $\lambda$ is highly relevant to $\Omega$, since the difference between our method and the GRWA becomes evident when $\Omega$ is large. Fig.~1 (d) shows the ground state energy as a function of the bias, from which one notes that our method gives an excellent result in comparison to the numerical exact one. In contrast to the GRWA, our improvement is definite.

Since the mean photon $\left\langle a^{\dagger}a\right\rangle$ and the correlation $\left\langle\sigma_{z}\left(a^{\dagger}+a\right)\right\rangle$
are simply depend only on $\lambda$, we can see the importance of the choice of $\lambda$ in detail. Quantitatively, we see that our calculation agrees with numerical exact result very well, while the GRWA presents incorrect result. Qualitatively, our method seize the important physics that the mean photon number and correlation are both dependent on $\Omega$ and $\epsilon$ (see Figs.~2 (c), (d) and (e), (f)), respectively. This feature is ignored by the GRWA. Our method is efficient on calculating most of the physics quantities, such as the average value of spin orientation (see Fig.~3).

\section{Applicable regime}
It would be valuable to briefly discuss the applicable regime of our method. The main idea of the GRWA is to consider the $\Omega$ term in the Hamiltonian as a perturbation. Thus, GRWA would work well when $\Omega/\omega$ is not large (see Fig.~4 (a)). If $\Omega$ is large, the performance of the GRWA will depend on the value of $g$. Here we discuss three cases. Case I: $g$ is large enough, the oscillator term and the coupling term together construct a displaced oscillator and they dominate. Hence the $\Omega$ term still can be considered as the perturbation and the GRWA works well in this case. Case II: $g$ is small enough, the $\Omega$ term is non-perturbative, and the displacement parameter $\lambda$ should be optimized. The $\lambda$ optimized GRWA is our variational method. Case III: $g$ is neither large nor small, we find that no matter how to optimize the $\lambda$, we can not lift the error between our method and the numerical exact one (see Fig.~4 (b)).  That means the form of the trial wave function in Eq.~\eqref{gs_wf} is no longer applicable. In fact, consider the completeness of the coherent state, the exact wave function is a linear combination of different coherent states and should take the form
\begin{equation}\label{comp}
|G\rangle_{exact}=\sum_{i=1}^{\infty}\left(\alpha_i\left|-_z,\lambda_i\right\rangle+\beta_i\left|+_z,-\lambda_i\right\rangle\right).
\end{equation}
where $i$ is the label to distinguish the coherent states with different eigenvalues.
Therefore, our method is the leading order approximation. When $\Omega/\omega$ is large and $g$ is comparable to $\sqrt{\omega\Omega}$, the higher order terms evidently involved and our scheme should be extended. Actually, in such regime, a novel quantum phase transition has been reported\cite{hwang2015,hwang2016,puebla2017,liu2017arxiv}, which reveals the competition of the coherent state components and indeed involves the higher orders in Eq.~\eqref{comp}.
\begin{figure}[!tb]
	\centering	
	\includegraphics[width=0.9\columnwidth]{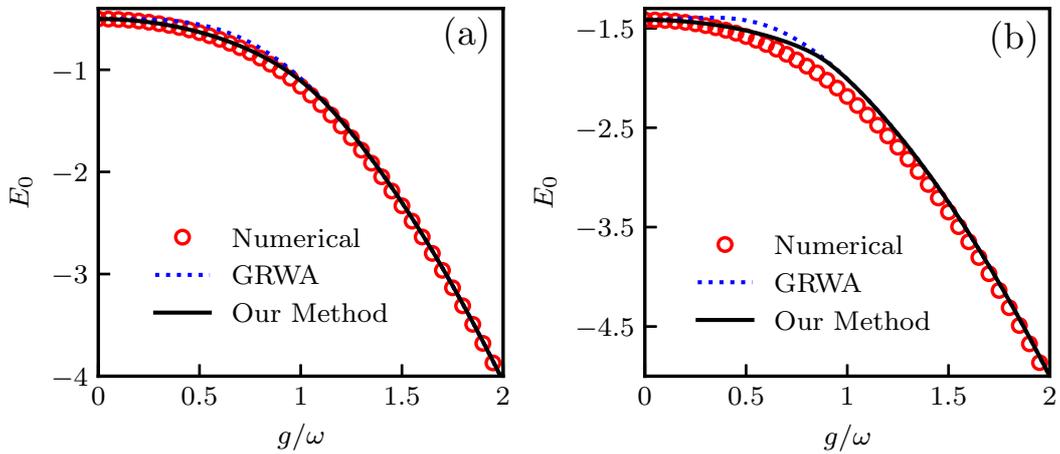}
	\caption{(color online)  The ground state energy of the system with different parameters, (a) $\epsilon=0.1$, $\Omega=1$, (b) $\epsilon=2$, $\Omega=2$.}\label{fig4}
\end{figure}
\section{Conclusion}\label{sec_4}
We have presented an analytical variational method for the biased quantum Rabi  model. The method improves the GRWA by further considering the variational principle. The key point is that we introduce a variational parameter $\lambda$, which is determined by minimizing the ground state energy.  Qualitatively, the mean photon number and the correlation of ground state are dependent on the frequency of the two-level atoms, and the bias strength in our method(on the contrary, the ground state photon number and correlation in GRWA are independent of the two-level atom and the bias, which is obviously not true). Quantitatively, comparing to the GRWA, we obtain a much accurate ground state energy. Our approach evidently improves the GRWA in the ultra-strong coupling regime. This improvement is extremely notable if the ratio of the $\omega/\Omega$ is small. The applicable regime of our method covers the ultra-strong coupling regime which has been achieved in the present experiment\cite{Fornd2017,Yoshihara2017,chen2017}. Therefore, we expect that this study will be helpful for understanding the relevant experiments.

\end{document}